# Possibility of using of the measured frequency $f$ instead of ω self-generated frequency

Manana Kachakhidze, Nino Kachakhidze-Murphy

**Abstract**

Possibility of using of the measured frequency $f$ instead of ω self-generated frequency in order to compute of earthquake magnitude is discussed.

**Introduction.**

At the present time we meet rather diverse and interesting papers, published in the scientific world on the basis of ground-based and satellite data of earth VLF/LF and ULF electromagnetic (EM) emissions observed in earthquake preparation period (Biagi, et al.,1999; Hayakawa, 1999; Hayakawa, et al.,1999; Uyeda, et al., 2000; Gershenzon and Bambakidis, 2001; Hayakawa and Molchanov, 2002; Hattori, et al., 2004;Bahat, et al., 2005; Freund, et al., 2006; Parrot, 2007; Eftaxias, et al., 2007a; 2007b; 2009; 2010; Biagi, et al., 2009; Pulinets, 2009; Rozhnoi, et al., 2009; Contadakis, et al., 2010; Papadopoulos, et al., 2010; Zlotnicki, et al., 2010; Biagi, et al., 2013; Hayakawa, et al., 2013; Ouzounov, et al., 2013).
These phenomena are detectable both at laboratory and geological scale (Hayakawa and Fujinawa, 1994; Hayakawa, 1999; Hayakawa, et al., 1999; Gershenzon and Bambakidis,
2001; Hayakawa and Molchanov, 2002; Bahat, et al., 2005; Eftaxias, et al., 2007a; Muto, et al., 2007; Hadjicontis, et al., 2007). Observations proved that when a material is strained, electromagnetic emissions in a wide frequency spectrum ranging from kHz to MHz are produced by opening cracks (Hayakawa and Fujinawa, 1994; Hayakawa, 1999; Hayakawa and Molchanov, 2002; Bahat, et al., 2005; Eftaxias, et al., 2007a; 2009).On the large (geological) scale, intense MHz and kHz EM emissions precede earthquakes that: (i) occurred in land (or near coast-line), (ii) were large (magnitude 6 or larger), or (iii) were shallow (Eftaxias, et al., 2002, 2004, 2006, 2007b; Kapiris, et al., 2004a; Karamanos, et al., 2006). More importantly, the MHz radiation precedes the kHz at geophysical scale (Eftaxias, et al., 2002; Kapiris, et al., 2004; Contoyiannis, et al., 2005).
The kHz EM anomalies were associated with the fracture of asperities that were distributed along the L'Aquila fault sustaining the system. The aspect of self-affine nature of faulting and fracture is widely documented from both, field observations and laboratory experiments, and studies of failure precursors on the small (laboratory) and large (earthquake) scale. (Eftaxias, et al, 2009).
Prof. Biagi in work „European Network for collecting VLF/LF radio signals (D5.1a)" pointed out that „ From several years, a research into the interaction between seismic activity and disturbances in radiobroadcasts has been carried out. One of the first results was obtained using 18 MHz receivers on the occasion of the great (M=8.5) Chilean earthquake of May 22, 1960, but it was published 22 years later. The receivers were part of a network used for studying cosmic noise.

Later, pre-seismic disturbances in VLF radio signals in the 20-60 kHz frequency band, have been presented mainly by Japanese and Russian researchers (Hayakawa et al., 1996; Hayakawa et al., 2002; Molchanov and Hayakawa, 1998).

At the same time, pre-seismic disturbances on LF (150-300 kHz) radio broadcasts were proposed mainly by Italian researchers (Bella et al, 1998; Biagi, 1999; Biagi et al., 2001a,b; Biagi and Hayakawa, 2002, Biagi et al., 2004). We can call all these disturbances *radio precursors*.

As it concerns the earthquakes selection, the three following possibilities have be adopted: a) earthquakes with Mw ≥ 5.0 located inside the 5$^{th}$ Fresnel zone of the different radio paths; b) earthquakes with Mw ≥ 5.0 occurred inside a circle with 200 km radius around each receiver; c) earthquakes with Mw ≥ 5.0 occurred inside a circle with 200 km radius around each transmitter. The rule a) takes into account several results which indicate that the area inside the 5th Fresnel zone is the most sensitive as for the seismic disturbances on the radio propagation (Molchanov and Hayakawa, 1998; Molchanov et al. 2006; Rozhnoi et al., 2005). The rules b) and c) are based on the dimension of the area interested by possible pre-seismic effects (Dobrovolsky et al., 1989; Kingsley et al., 2001, Biagi et al.,2013).

**Discussion and results.**

It is known that VLF / LF Radio Networks of different countries fix $f$ frequency of electromagnetic emissions. Accordingly, in every scientific work, concerns the electromagnetic emissions at the prior of the earthquake, the same $f$ frequency is considered (Eftaxias et al., 2000, Biagi et al., 2013; Hayakawa et al., 2013; Ouzounov et al., 2013).

Due to the fact that electromagnetic emissions was fixed vividly in a seismic zone during foreshocks and aftershocks, a models, based on electrodynamics, was created for its explaining. Besides, it was obvious that electromagnetic emissions occurred in a definite segment of the earth crust, at the impact of external tectonic stress. Thus, free oscillations should not be present here and we had to take into consideration the fact that in this case we had to deal with vibrating, distributed system. Alongside with it, it is apparent that with the geological point of view, fault length in the focus is increased as a result of impact of tectonic stress.

Therefore, we had to find a means which would communicate any parameter characteristic for electromagnetic oscillations generated by the increase of tectonic stress with any parameter characteristic for mechanical oscillations conditioned by the increase of the very tectonic stress.

Studies showed that such two parameters are frequency fixed at electromagnetic emission and length of the fault created in earthquake focus (Eftaxias, et al., 2002, 2004, 2006, 2007b; Kapiris, et al., 2004a; Karamanos, et al., 2006, Bella et al, 1998; Biagi, 1999; Biagi et al., 2001a,b; Biagi and Hayakawa, 2002, Biagi et al., 2004, Biagi et al.,2013, Mjachkin, 1975; Ulomov, 1993).

It is known that in the case of free electromagnetic oscillations which take place in a real oscillating circuit which is presented as a series connection of the inductance coil L, capacitor C and electrical resistance-R :

$$f = \omega/2\pi, \qquad \omega^2 = \frac{1}{LC} - \frac{R^2}{4L^2}$$

Medium where an earthquake is prepared is a distributed system, where free electromagnetic oscillations don't take place. In this system tectonic stress permanently acts, which contributes to alteration of physical and chemical properties of the environment. This is why relation of elements positions (or element groups) play substantial role with the view of system functioning.
It is known that distributed systems are characterized by distribution of functions, of resources between multiple elements (nodes) and by the absence of integrated control center. In distributed systems, when it is impossible to separate a definite point, each canal of energy transmittance is considered a pair of poles (points of connection).
Thus, in this environment, free oscillations, neither mechanical nor electromagnetic, can't and don't exist.
More than that, earthquake preparation area is the so-called area of parametric impact, since at the impact of external (tectonics) forces at least one parameter mass, flexibility (mechanical systems), capacity or inductance (electronic systems) is changed. In addition, this impact to a certain extent, is periodical.
With the seismologic point of view this implies that at the every seismically active regions tectonic stress permanently acts, but to a definite stage it doesn't result in significant changes in the earth crust until it approaches the limit of geological strength of the environment (this process depends on geological structure of the environment). But a moment comes when definite rocks can't resist the created stress and cracks start to appear. This process is well known for seismologists and is perfectly described by avalanche-instable model (Mjachkin, 1975).
Decrease of tectonic energy that is accumulated in seismic zone proceeds at the expense of cracks formation, but due to the fact that tectonic stress permanently acts in the region, in the following periods tectonic energy again reaches the limit of geological strength of the medium and again new cracks are formed. Thus, tectonic stress, with the view of impact on rocks, is characterized by definite periodicity.
This process is progressing till the main shock occurs, when the major portion of the accumulated tectonic energy is released.
Now let's consider the model offered by us (virtual contour) and for simplicity assume that we are considering a system with a degree of freedom equaling to one.
It is known that increase of tectonic stress (avalanche- unstable model) incites formation of the main fault, which, in case of large earthquake can reach 300 km in length (e.g. Venchuan earthquake). Since this process takes place in the rocks, which is a distributed system, increase of the fault length implies that at the impact of external tectonic stress the collection of elements (nodes) is multiplied and respectively, the capacity (or inductance) of certain definite element of a virtual contour offered by us is increased. For simplicity, we will consider changes of only one parameter – capacity (the same should be said about the second parameter – inductivity). We'll confine ourselves to consideration of a case when periodical changes, that is, periodical modulations of one parameter take place.
This process of generation of oscillations in seismogenic area, which is conditioned by periodical changes of power-consuming parameter of oscillation system (in our case capacity or inductivity) is a typical example of parametric resonance of oscillations (Migulin,1978).
It should be stated that the value of changes of oscillation energy in the contour at the impact of parameters is commensurable to the energy accumulated in the system.
In addition, the relation between the parameter changes and oscillations generated by them has the following form:
$$p = \frac{2f}{n}$$

where n- is number of parameter changes in the period of excited oscillations, n=1,2,...; $p$ - frequency of parameter changes, and $f$ - is the frequency of excited oscillations (Migulin, 1978). Increase of an amplitude of excited oscillation, and correspondingly, increase of the system energy takes place at the expense of external forces, which results in alteration of parameters too. Of course parametric excitation of oscillations is possible in case of alteration of one of the power-consuming parameters (C or L). Alteration of R can lead only to changes in dissipation law, while the system still remains dissipative.

Besides, there are data in special literature stating that in the earthquake preparation zone drastic increase in electric conductivity takes place. (Morozova, et al., 1999; Goto, et al., 2005; Kovtun, 2009).

We have already stated above that the phenomenon when the progressing oscillation process is formed in the contour the frequency of which strictly depends on the frequency of external parametric impact and caused by the very impact is parametric resonance of oscillation

Parametric resonance takes place when definite ratio is fulfilled between the parameter changes of $p$ frequency and $f$ frequency of excited (generated) oscillations, which are very close to $\omega$ self-generated frequency of the excited system ($p=2f/n$).

Thus, it is quite acceptable to use the measured frequency $f$ instead of $\omega$ self-generated frequency during computations (Kachakhidze et al., 2015).

$$f \approx \omega = k\frac{c}{l} \quad (1)$$

where $l$ – is the length of the fault in the focus.

Therefore, when we need to use formulae (1) in practice, we use result getting for $\omega$ self-generated frequency (Kachakhidze et al., 2015).

It is clear that in case of parametric resonance in non-linear systems the presence of non-linearity alongside with the increase of amplitudes will result in changes of oscillations self - frequencies. Respectively, the term of resonance will be violated and the amplitudes of parametrically excited oscillations will be restricted.

In case of a contour of non-linear inductivity, at periodical alteration of $L$ and $C$, a formula describing oscillations taking place in the system will have no simple form. In this case we can obtain only approximate solutions. Besides, we have to consider degree of non-linearity of the system, since non-linearity element of the system has the limiting role, that enables us to obtain final, limited amplitudes of parametrically excited oscillations in conservative system. Anisochronous of the system leads us to the fact that in case of parametric excitation of oscillations, self-frequency of a system is changed together with the increase of the amplitude and the system passes to the corresponding parametric excitation zone. This results in the decrease of energy delivered to the system, which, in its turn, changes the parameter and because of this the amplitude growth is restricted.

We not consider in details a parametric excitation theory for distributed conservative systems.

In our model we use only one of the results of parametric excitation theory in vibrating systems, stating that at definite conditions frequency of excited oscillations approaches self - excitations of the excited system.

It is significant to note that in case of parametric excitation the system (focus area) stays in the undisturbed state ($f$ frequency oscillations are not excited in it), for $f$ all meanings, when $f<f_2$ and $f >f_1$. But for $f$ all meanings which is inside of ($f_1$-$f_2$) frequency diapason, the wavy processes with the final A amplitude will arise. The value of amplitude significantly depends on degree of nonlinearity of the system ($\gamma$ value). Because of dependence on nonlinearity character ($\gamma$), oscillations fall up to zero or increase outside the limits of diapason ($f_1$-$f_2$) (the first area of parametric excitation $2\omega/p = \omega/f \approx 1$), while the system passes to stable, undisturbed state in case of these $f$ values (Fig. 1) (Migulin, 1978).

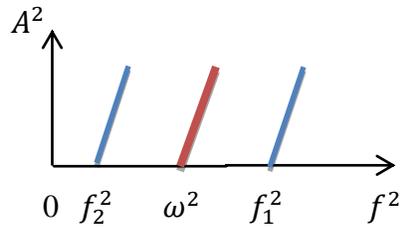

Fig. 1

As to the reason why our model is based on frequency values and not on any parameter: this is conditioned by theory of the parametric resonance arising in the vibrating system which is proved by laboratory and field observations: in the periods of earthquake preparation, earthquake occurring and following periods (up to full release of tectonic stress), it is namely the frequency parameter that extremely adequately expresses the processes going on in the focus (for $M \geq 5$ earthquakes) which is characterized by absolutely definite character changes: e.g. during all large earthquakes, during a definite period preceding the earthquake, electromagnetic emission frequency in MHz range is changed, while immediately before earthquake it passes to kHz (Eftaxias, et al., 2009; Biagi et al.,2013; Hayakawa et al., 2002;). More than that, in the special scientific literature it was explicitly fixed that an important feature, observed both on a laboratory and a geological scale, is that the MHz radiation precedes the kHz one (Eftaxias et al., 2002 and references therein).

It should be emphasized that the dependence between electromagnetic emission frequency and fault length in the earthquake focus (1) reflects by rather high precision the processes of foreshocks, main shock and aftershocks.

"Short-term earthquake (EQ) prediction is considered the highest priority for social demands, and represents the most important in term of possible alerts to be issued to prevent casualties. However, short-term EQ forecasting represents also a big scientific challenge, as demonstrated by the debate that is still open between those who are in favour and those who are against the possibility to predict earthquakes in a reasonably short time.

However, it is in the opinion of the authors of the present deliverable that the scientific level of the studies is still not satisfactory, and that they need to be promoted and supported, not to get short-term operational answers (impossible to get at the present), but to prepare the necessary background (in terms of observations, experiments, methodologies, theories, models) as large as possible in order to solve the problem of earthquake prediction in the future". (Telesca, et.al.2013).

**Conclusion:** We use the results of parametric excitation theory in vibrating systems, stating that at definite conditions frequency of excited (generated) oscillations approaches self - excitations of the excited system.